\documentclass[%
 reprint,
 english,
 aps,
 amsmath, amssymb 
]{revtex4-2}

\usepackage{babel} 
\usepackage[utf8]{inputenc} 

\usepackage{wrapfig}
\usepackage{booktabs}
\usepackage{dcolumn}
\usepackage{tabularx, array, rotating}
\usepackage{makecell}
\usepackage{multirow}
\usepackage{float}

\usepackage{mathtools, amsthm} 
\usepackage{bm}
\usepackage{mathrsfs}

\usepackage{tikz, tikz-cd}
\usepackage{adjustbox}

\usepackage{enumitem}
\usepackage{xcolor}
\usepackage{hyperref}

\DeclareMathOperator{\dvg}{div}

\DeclareMathOperator{\curl}{curl}

\newcommand{\bR}{\mathbb{R}}

\newcommand{\Hdiv}{H^{\dvg}}
\newcommand{\Hdivzero}{H^{\dvg 0}}

\newcommand{\Hcurl}{H^{\curl}}
\newcommand{\Hcurlzero}{H^{\curl 0}}
\newcommand{\Noslipspace}{H^{\curl,\dvg 0}_0}
\newcommand{\Bspace}{H^{\curl,\dvg 0}}
\newcommand{\Hone}{H^1}

\newcommand{\pprofile}{\mathrm{p}}
\newcommand{\Fprofile}{\mathrm{F}}
\newcommand{\iotaprofile}{\mathrm{iota}}

\newtheorem{theorem}{Theorem}

\newtheorem{definition}[theorem]{Definition}

\begin{document}

\title{Computational boundary specification in 3D fixed-boundary\\magnetohydrodynamic equilibrium modeling}

\author{Alan A. Kaptanoglu}
\affiliation{%
 Courant Institute of Mathematical Sciences, New York University, 251 Mercer St, New York, 10012, NY, USA
}%
\author{Tobias Blickhan}
\affiliation{%
 Courant Institute of Mathematical Sciences, New York University, 251 Mercer St, New York, 10012, NY, USA
}%
\begin{abstract}
Outside the core of the plasma, the plasma current and pressure rapidly transition to zero in a scrape-off or edge region or plasma-vacuum interface. However, existing tools for fixed-boundary magnetohydrodynamic equilibria in 2D and 3D domains $\Omega$ typically prescribe the computational boundary $\partial\Omega$ interior to this transition layer. We (1) argue that a more realistic and robust assumption is to define the computational boundary exterior to this transition layer, in a vacuum-like region where $J|_{\partial\Omega} \sim p|_{\partial\Omega} \sim 0$, (2) show that, without this boundary change, existing coil optimization routines for 3D toroidal equilibria (stellarators) should be changed to match free-boundary equilibrium requirements, and (3) derive an algorithm for a fixed-boundary 3D equilibrium solver compatible with a very general computational boundary, with conditions $B \cdot n|_{\partial\Omega} \neq 0$ (not necessarily a flux surface), $p|_{\partial\Omega} \neq \text{const.}$ (not necessarily an isobar), and $J \times n|_{\partial\Omega} \neq 0$.

\end{abstract}
\maketitle

\section{Introduction}\label{introduction}
\begin{figure}[H]
\centering
\includegraphics[width=0.9\linewidth]{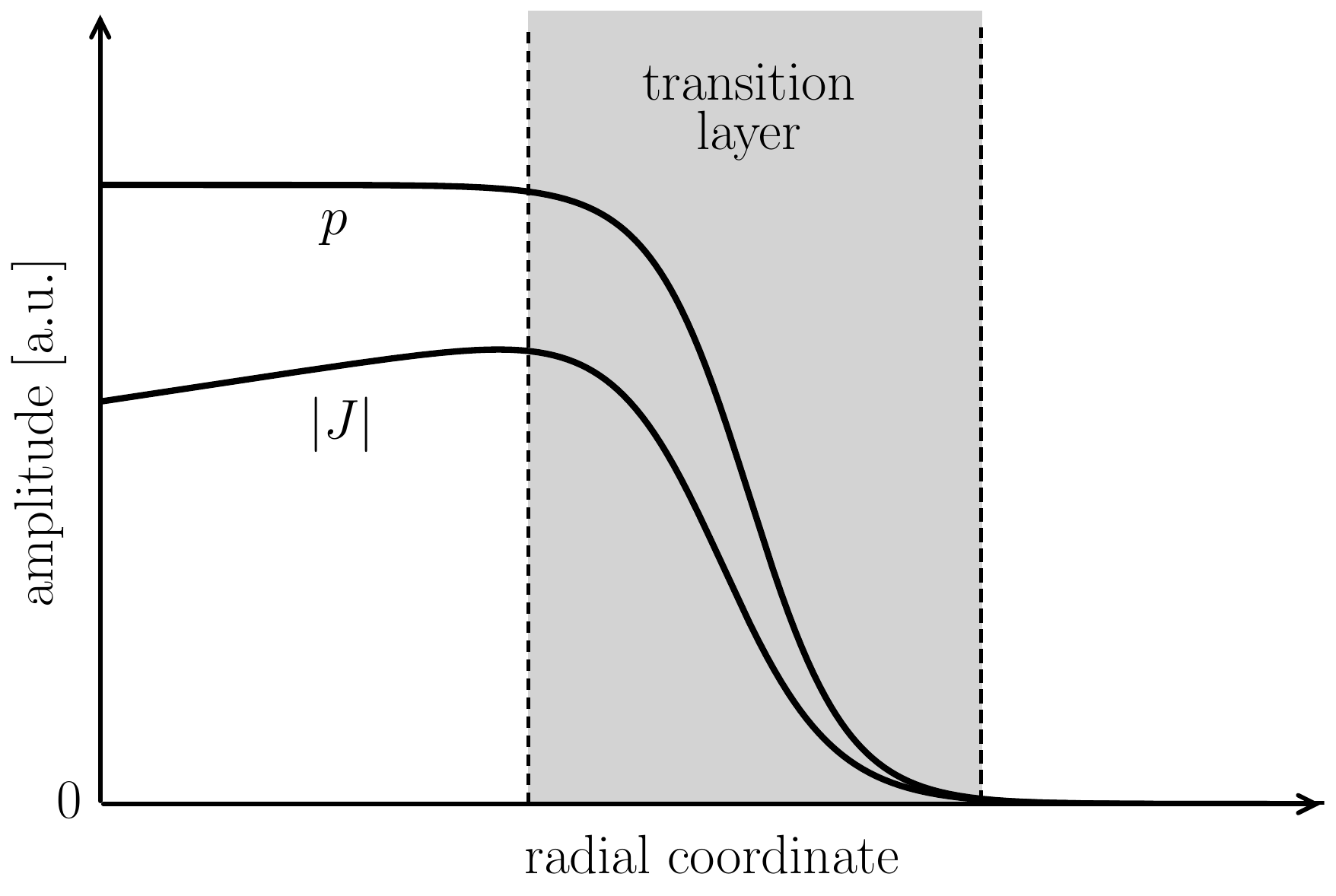}
\caption{Representative $p$ and $J$ profiles. $J|_{\partial\Omega}$ and $p|_{\partial\Omega}$ approximately vanish on the outer surface.}
\label{fig:boundarychoices}
\end{figure}

\begin{figure}
    \centering
    \includegraphics[trim=3cm 8cm 3cm 6cm, clip, width=\linewidth]{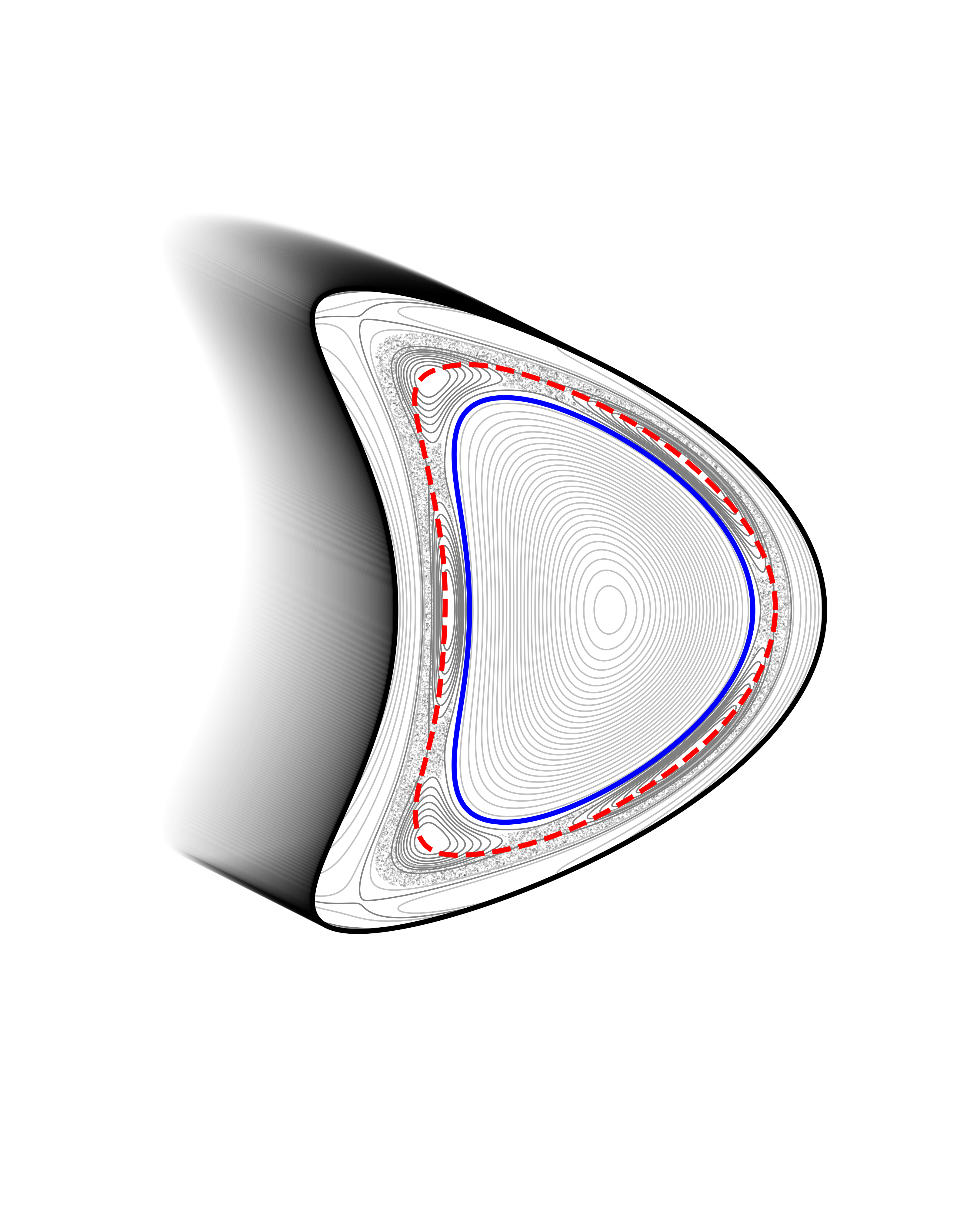}
    \caption{
    Different choices of computational boundary. \textbf{\textcolor{blue}{Blue}}: traditional closed-flux-surface boundary $\partial\Omega$.
    \textbf{\textcolor{red}{Red}}: one possible isobar drawn through a chaotic/island
    region. \textbf{Black}: Computational boundary in the vacuum region where $p|_{\partial\Omega} = J|_{\partial\Omega} = 0$.
    }
    \label{fig:CBS}
\end{figure}

Fixed-boundary magnetohydrodynamic (MHD) equilibrium calculations depend strongly on where the computational boundary (sometimes called the ``plasma boundary'', which we avoid here specifically because it need not be) is placed, relative to the typical plasma core-edge transition layer or plasma-vacuum transition layer. 
Figure~\ref{fig:boundarychoices} shows typical profiles for the current density $J$ and pressure $p$ for a fusion-relevant plasma in a toroidal domain. Along with Fig.~\ref{fig:CBS}, it illustrates that there is a boundary-placement issue, where the appropriate MHD equilibrium calculation assumptions on the profiles of $J$ and $p$ depend on where the boundary is placed. This computational boundary specification (relatedly, also the specification of the pressure and current on a given boundary) will henceforth in this paper be abbreviated as the CBS. For concreteness, we will consider only \textit{static} (zero equilibrium flow) equilibria throughout this manuscript.

\subsection{Fixed-boundary 2D MHD equilibria}\label{sec:2D_equilibria}
In fixed-boundary 2D axisymmetric geometry, MHD equilibrium is described by the Grad-Shafranov equation, where the CBS is entirely controlled by specifying two free profiles, e.g. the pressure $\pprofile$ and $\Fprofile$, as a function of a flux function $\psi$. We will show this briefly. Throughout this work, we denote these \emph{profile functions} with upright Roman fonts, e.g. $\pprofile : [0,1] \to \bR$, to differentiate them from the flux functions $p: \Omega \to \bR$, where $\pprofile(\psi) = p$. 

\paragraph{Grad-Shafranov equation:}
Let $\Omega$ be a solid torus, not necessarily with circular cross-section, with $Z$-axis of symmetry, denote the angle of revolution around the axis by $\phi$ (the toroidal angle), and by $R$ the distance to it. Then, 
\begin{equation}
B = \frac{1}{R}\nabla\psi \times e_{\phi} + \frac{\Fprofile(\psi)}{R} e_{\phi}, \quad B\cdot\nabla\psi = 0 \quad \text{ in } \Omega,
\end{equation}
and the fixed-boundary problem is: given a smooth, simply-connected plasma cross-section
$\Omega \cap \{ \phi = 0 \} \subset \mathbb{R}^2$ with boundary curve $\partial\Omega \cap \{ \phi = 0 \}$, satisfying $B\cdot n = 0$ on $\partial \Omega$
and corresponding to $\psi = \psi_b$, and given profile data $\pprofile$ and
$\Fprofile$, find $\psi$ such that
\begin{align}
\Delta^* \psi &= - R^2 \pprofile'(\psi) - \Fprofile(\psi)\Fprofile'(\psi) &\text{in } \Omega \cap \{ \phi = 0 \},
\\
\psi &= \psi_b &\text{on } \partial\Omega \cap \{ \phi = 0 \}, \notag
\end{align}
where
\begin{align}
\Delta^* \psi := R\frac{\partial}{\partial R}\left(\frac{1}{R}
\frac{\partial \psi}{\partial R}\right) + \frac{\partial^2 \psi}{\partial Z^2}.
\end{align}
This is the classical Grad--Shafranov reduction of static
axisymmetric ideal-MHD equilibria
\cite{grad1958,shafranov1966,freidberg1987}. We have taken $\mu_0 = 1$ throughout this work. 

\paragraph{Boundary values:}
The current density on the boundary can be explicitly written as 
\begin{align}
J = \Fprofile'(\psi_b) B
+ R \, \pprofile'(\psi_b) e_{\phi} \quad \text{on } \partial\Omega \cap \{ \phi = 0 \}.\label{eq:JedgeGS}
\end{align}
Hence $J\cdot n=0$ on the boundary holds in the axisymmetric case, but the tangential component $J_t$ need not vanish
there unless additional edge conditions such as $\pprofile'(\psi_b)=0$ and
$\Fprofile'(\psi_b)=0$ are imposed~\cite{guazzotto2021part1,guazzotto2021part2,fitzpatrick2024inverse}. 
Indeed, many analytical and numerical Grad-Shafranov solutions exhibit finite tangential current on the computational boundary, corresponding to a CBS where the computational boundary is chosen on the interior of the transition layer~\cite{solovev1968,cerfon2010,guazzotto2007,pataki2013,xu2019vacuum}. 

Arguably, this is a sensible choice for 2D problems where a flux function is well-defined everywhere and nested flux surfaces generally exist in the plasma core (X-points and other critical points of $\psi$ are possible of course), while the transition layer and the edge dynamics in fusion devices may exhibit magnetic islands or chaos, and in some regimes may not be well-described as magnetohydrodynamic equilibria anyways. 

\paragraph{Impact on inverse problems and control:}
Moreover, experimental equilibrium reconstruction commonly treats a discharge as a time-indexed sequence of static equilibria: EFIT-type and LIUQE-type reconstructions solve an inverse Grad--Shafranov equilibrium problem from diagnostic data at each time slice, and real-time variants compute these quasi-equilibria quickly enough for feedback control \cite{lao1985efit,lao2005diiid,ferron1998realtime,moret2015liuqe}.  This quasi-static viewpoint is sensible when the evolution is slow compared with the MHD equilibration time, but it also makes the boundary specification important.  If the edge is evolving or exhibiting 3D effects, the free functions (e.g. pressure and toroidal current) at the computational boundary $\partial \Omega$ can change; moreover, it is not even clear that $\partial\Omega$ remains a flux surface, so the boundary itself may need to be redefined.  

\paragraph{Self-consistency:}
In stellarator modeling, self-consistency between transport, bootstrap current, and equilibrium is often obtained by iterating between a 3D MHD equilibrium calculation and neoclassical or transport calculations, while penalizing inconsistency e.g. between the equilibrium current profile and a predicted bootstrap current~\cite{landreman2014sfincs,landreman2022selfconsistent,bader2020qhs}.  Thus the pressure, current profiles and, in edge-sensitive situations, the boundary condition and boundary location, may all change when the equilibrium is recomputed.  This calls into question the utility of a fixed-boundary equilibrium calculation whose CBS is not explicit.

\paragraph{Absence of nested flux surfaces:}
The problem is even more pronounced in 3D toroidal equilibria (stellarators) or even 3D toroidal \textit{vacuum} solutions; generically such equilibria will exhibit no flux surfaces at all~\cite{reiman1986,hudson2002,kanno2002,suzuki2006,hudson2012,enciso2025jems}!
Moreover, in 3D stellarator geometries, magnetic islands and chaos can show up generally, even in the plasma core, and stellarators operate quite close to steady-state. Complex 3D coils must also be found to create the magnetic fields, and often use plasma boundaries determined from fixed-boundary calculations. The motivation to keep the transition layer and edge regions out of the magnetohydrodynamic equilibrium modeling is much less clear in this setting. 

\subsection{Choice of function spaces}\label{sec:function_spaces}
We will use the Hilbert space of square-integrable functions on this domain, $L^2(\Omega; \mathbb{R}^3)$.
We equip all of the following function spaces with the standard inner product; for any $u, v \in L^2(\Omega; \mathbb{R}^3)$,
\begin{align}
\label{eq:inner_product_definition}
    ( u, v)_{L_2(\Omega)} = \int_\Omega u \cdot v \, dV.
\end{align}
All differential operators below are understood weakly. The domain $\Omega \in \mathbb{R}^3$ is bounded with $C^{1,1}$ boundary (i.e. locally the graph of a function with Lipschitz-continuous derivative). 
\begin{definition}[Function spaces]
    The spaces of vector fields on $\Omega$ with weak grad, curl and divergence are defined as:
    \begin{align}
        \Hone(\Omega; \bR)    &:= \{ p \in L^2(\Omega; \bR): \nabla p \in L^2(\Omega; \bR^3) \},  \\\notag
        \Hone(\Omega; \bR^3)    &:= \{ v \in L^2(\Omega; \bR^3): \nabla v \in [L^2(\Omega; \bR^3)]^3 \},  \\\notag
        \Hcurl(\Omega; \bR^3) &:= \{ E \in L^2(\Omega; \bR^3): \nabla\times E \in L^2(\Omega; \bR^3) \},  \\\notag
        \Hdiv(\Omega; \bR^3)  &:= \{ B \in L^2(\Omega; \bR^3): \nabla\cdot B \in L^2(\Omega; \bR) \}.  
    \end{align}
    When no ambiguity arises, we will suppress the domain and/or codomain in the notation. The homogeneous trace subspaces used in the paper are:
    \begin{align}
        \Hone_0(\Omega; \bR)    &:= \{ p \in \Hone(\Omega; \bR): p|_{\partial \Omega} = 0 \}, 
        \\\notag
        \Hone_0(\Omega; \bR^3)    &:= \{ v \in \Hone(\Omega; \bR^3): v|_{\partial \Omega} = 0 \}, \\\notag
        \Hcurl_0(\Omega; \bR^3) &:= \{ E \in \Hcurl(\Omega; \bR^3): E \times n|_{\partial \Omega} = 0 \},   \\\notag
        \Hdiv_0(\Omega; \bR^3)  &:= \{ B \in \Hdiv(\Omega; \bR^3): B \cdot n|_{\partial \Omega} = 0 \}.
    \end{align}
Lastly, we introduce the spaces
\begin{align}
    \Hcurlzero(\Omega) &:= \{ E \in \Hcurl(\Omega) : \nabla \times E = 0 \}, \\
    \Hdivzero(\Omega) &:= \{ B \in \Hdiv(\Omega) : \nabla \cdot B = 0 \}, \notag \\
    \Bspace(\Omega) &:= \{ B \in (\Hcurl \cap \Hdiv)(\Omega): \nabla \cdot B = 0 \}. \notag
\end{align}
The spaces $\Hcurlzero_0(\Omega), \Hdivzero_0(\Omega)$ and $\Noslipspace(\Omega)$ are defined analogously with their respective boundary conditions. It holds that $(\Hcurl_0 \cap \Hdiv_0)(\Omega; \bR^3) = H^1_0(\Omega; \bR^3)$.
The space $\Noslipspace$ consists of solenoidal fields with no-slip velocity boundary conditions $v|_{\partial\Omega} = 0$.

For the force-balance residual we will use the following $L^2(\Omega)$ force class,
\medmuskip=0mu
\thickmuskip=2mu
\begin{align}
    \mathcal F_2 := \{(B, p): B\in \Bspace,\; p\in\Hone,\; J\times B\in \Hdiv\}.
\end{align}
\thickmuskip=4mu
\medmuskip=4mu
Thus $J \times B - \nabla p$ is an $L^2$ residual whenever $(B, p)\in\mathcal F_2$. This is the minimum regularity required to define the residual $\| J \times B - \nabla p \|^2_{L^2(\Omega)}$, but our method is applicable to a wider class of functions (c.f. Appendix~\ref{sec:energy_decrease}).
\end{definition}

\subsection{3D fixed-boundary MHD equilibria}\label{sec:3D_equilibria}
Fixed-boundary, three-dimensional, static, ideal  magnetohydrodynamic equilibrium in a toroidal volume $\Omega$ is given by the force balance equation coupled with the Maxwell equations.
This can be formulated as a search for $(B,p) \in\mathcal F_2$ such that
\begin{align}
\label{eq:equilibrium}
     J \times  B &= \nabla p, \,\,\, \nabla\cdot B = 0,  \,\,\,\nabla\times  B = J  \text{ in }\Omega, \\   \notag B\cdot n &= 0 \text{ on }\partial\Omega, 
     \quad \oint_{\Omega \cap \{\phi = 0\}} B\cdot n\, dS = \Phi_0.
\end{align}
Here, $\Omega \subset \mathbb{R}^3$ denotes a bounded $C^{1,1}$ solid non-axisymmetric torus, $ n$ the unit normal on $\partial \Omega$ and a user-supplied constant $\Phi_0$ fixes the device scale (the harmonic part of the solution) by specifying the total toroidal flux through the poloidal cross-section $\Omega \cap \{\phi = 0\}$.

\paragraph{Profile specification:}
Most codes require that profiles such as the spatially-varying pressure $\pprofile$ and rotational transform, $\iotaprofile$, must be provided. 
Stellarators require that the full 3D equations must be solved in a toroidal volume $\Omega$ where nested flux surfaces are not guaranteed.
Nonetheless, most stellarator equilibrium codes, such as the commonly used VMEC~\cite{hirshman1983}, VMEC++~\cite{schilling2025numerics}, GVEC~\cite{Hindenlang_2025}, and DESC~\cite{dudt2020desc,panici2023} codes (see \citep[Table 1]{blickhan2025}, for a comprehensive overview) assume that nested magnetic flux surfaces exist and therefore the surfaces can be labeled by their corresponding value of the toroidal magnetic flux $\psi$. This implies also that $p = \pprofile(\psi)$ and $\iota = \iotaprofile(\psi)$ are flux functions. Nested flux surfaces are often a useful assumption because it simplifies the problem and we expect that stellarators with good confinement should generally exhibit large volumes of nested flux surfaces. 
Unfortunately, this assumption also leads to singularities that can be found even in weak form solutions~\cite{loizu2015magnetic,lazerson2016verification} and can generally prevent nested flux surface codes from finding solutions with volume-averaged force balance beyond some tolerance~\cite{imbert2024introduction}. 

Regardless of whether the solver is nested or non-nested, essentially all solvers formulate the fixed-boundary as solving the force balance equations~\eqref{eq:equilibrium}, given a flux surface $\partial \Omega$ and at least \textit{initial} profiles such as $\pprofile$ and $\iotaprofile$ as a function of a flux coordinate.
For instance, in non-nested codes like SPEC, the pressure profile is provided as a piecewise constant function across computational surfaces used for solving the interior problem~\cite{hudson2020specfb,baillod2021,hudson2025}. SIESTA inherits a great
deal of structure, including profiles from the VMEC initial condition~\cite{hirshman2011,peraza2017fb,peraza2017}. PIES accepts profile information as initial conditions and then iterates the fields to a non-nested force-balanced state that may contain islands or stochastic regions~\cite{reiman1986,hudson2002}.
Lastly, the extent of the fixed-boundary functionality in HINT it is not clear to us but they also appear to solve some problems where the initial profiles are inherited~\cite{suzuki2017hint}.

\subsection{Conditions for force balance at the boundary}\label{sec:force_balance_on_boundary}

\paragraph{Edge pressure and current:}
As in 2D, a 3D boundary with $p|_{\partial\Omega}=p_0$ constant and vanishing edge current can be encouraged in nested-surface codes by choosing pressure and current profiles that flatten near $\partial\Omega$. On a flux-surface boundary, force balance and $p|_{\partial\Omega}=p_0$ imply formally that $J\cdot n=0$ away from magnetic-field nulls: $(J \times B) \times n = \nabla p \times n$ holds due to force balance and $(J \times B) \times n = (B \cdot n) J - (J \cdot n) B$. If we now assume sufficient regularity to evaluate this relation on $\partial \Omega$, then $B  \cdot n |_{\partial \Omega} = 0$ there and, when $p$ is constant on the boundary, $\nabla p \times n |_{\partial \Omega} = 0$. We arrive at $((J \cdot n) B) |_{\partial \Omega} = 0$ so either $B |_{\partial \Omega}$ vanishes identically or $(J \cdot n) |_{\partial \Omega}$ does.
If, in addition, $\nabla p \cdot n|_{\partial \Omega} = 0$ on the boundary, then $J \times B|_{\partial \Omega} = 0$ there and $J|_{\partial \Omega}=\lambda B|_{\partial \Omega}$ on non-null field lines. If the edge-current profile is flattened in a neighborhood of the boundary, for example by taking the relevant toroidal-current function constant on $\psi \in[\psi_b-\delta,\psi_b]$ (with the orientation of $\psi$ chosen so that this interval lies inside the plasma), then $\lambda=0$ on $\partial\Omega$.

We note here that assuming that $J \times B$ and $\nabla p$ admit tangential traces is not a problem as these conditions do not play the role of boundary conditions, rather they are consequences of the force equilibrium and hence the target of our iterative algorithm, not boundary conditions enforced throughout iterations (as, for example, $B \cdot n|_{\partial \Omega} = 0$). If desired, they can be enforced using a penalty method.

\paragraph{Incompressible variations and pressure as a Lagrange multiplier:}
Regarding the specification of profile information, our recent code MRX~\cite{blickhan2025} is an exception for finite-$\beta$ 3D toroidal equilibrium solvers; while it is typically initialized by a nested flux surface equilibrium (or any other appropriate toroidal coordinate system), and inherits the initial magnetic field, it does not accept profile information because the pressure is determined by an elliptic solve every iteration, i.e.,
\begin{align}
\label{eq:genericpelliptic}
\Delta p^{k+1} &= \nabla\cdot(J^{k+1} \times B^{k+1})
\quad\text{in }\Omega,
\\ \notag
\partial_n p^{k+1}&=(J^{k+1}\times B^{k+1})\cdot n \quad \text{on }\partial\Omega.
\end{align}
which enforces that $p$ plays the role of the gradient term in the Hodge decomposition of the Lorentz force $J \times B$. This guarantees that under admissible variations~\cite{chodura_3d_1981,park1986}, a minimum of the energy will satisfy force balance. However, alone it implies the pressure values can vary on the computational boundary; on a flux surface, the pressure must be constant, since $B\cdot\nabla p = 0$, \textit{if} the fields are in force balance. Therefore, in isolation this update need not produce a constant pressure on the boundary, but it will in practice converge to an isobar because the overall iteration converges to an equilibrium in force balance.

Note also that this problem is solved weakly.  The pressure need not be continuously differentiable, and it generally will not be if the magnetic field exhibits islands and chaos. This approach requires $J \times B \in \Hdiv(\Omega)$ to obtain $p\in H^1(\Omega)$ and $J\times B-\nabla p \in\Hdivzero_0(\Omega)$.

\paragraph{Boundaries that are not flux surfaces:}
We will make the case in this manuscript for specifying a more general computational boundary in fixed-boundary MHD equilibrium calculations. 
Towards this, suppose that we prescribed a constant-pressure surface (an isobar) where $J\times B|_{\partial\Omega} = \nabla p|_{\partial\Omega}$ should be satisfied but otherwise impose no other conditions on the fields on this surface.
Since the surface is an isobar, $\nabla p|_{\partial \Omega} = n \, \partial_n p|_{\partial \Omega}$. 
%
Dotting both sides of the force balance equation by $B$ and under sufficient regularity, it follows that,
\begin{align}
    0 = B \cdot \nabla p = (B \cdot n)\partial_n p \quad  \text{on }\partial\Omega.
\end{align}
We can see that if $\partial\Omega$ is not a flux surface, then $\partial_n p$ must vanish wherever the normal magnetic trace is nonzero and the fields are in equilibrium. 


\subsection{Free-boundary equilibrium calculations}

In a free-boundary problem the surface is no longer prescribed in
advance. Instead one solves simultaneously for the plasma equilibrium,
the vacuum field, and the interface $\partial\Omega$:
\medmuskip=0mu
\begin{align}
&J_p\times B = \nabla p, &\nabla\times B &= J_p, &\nabla\cdot B &= 0 &\text{in } &\Omega, \\
& &\nabla\times B_v &= 0, &\nabla\cdot B_v &= 0
&\text{in } &\Omega_v. \notag
\end{align}
\medmuskip=4mu

Here $B$ and $B_v$ denote the traces of the \emph{total field} on the
plasma and vacuum sides of the same candidate interface. Note that $B \neq B+B_v$ within one domain, but rather $B \big |_{\Omega \cup \Omega_v} = (B \oplus B_v)\big |_{\Omega \cup \Omega_v}$.

\paragraph{Interface conditions:}
For free-boundary calculations it is useful to write the interface conditions as residuals on a candidate surface $\partial\Omega$:
\begin{align}
\label{eq:freeboundaryresiduals}
R_n(\partial\Omega) &:= (B-B_v)\cdot n, \\ \notag
R_t(\partial\Omega) &:= n\times(B_v-B)- K, \\ \notag
R_T(\partial\Omega) &:= n\cdot(T_v-T_p)\cdot n,
\end{align}
where $K$ is a sheet current and $T_v$ and $T_p$ denote the respective Maxwell stress tensors. A self-consistent free-boundary equilibrium requires $R_n=R_t=R_T=0$ on the actual interface. The $R_T$ condition requires more regularity for $B$ \citep{arnold_finite_2010} and we will avoid such assumptions since free-boundary is not the focus in this work. For the important ``physical'' case where pressure goes to zero at the edge and the coils are well matched to the computational boundary, there is no explicit need for a sheet current~\cite{conlin2024}.

\subsection{The case for a vacuum-like CBS}
We make the case in this manuscript that a more natural and robust choice in 3D MHD equilibrium modeling is to choose a computational boundary $\partial \Omega$ and CBS where $J|_{\partial\Omega} = 0$ and $p|_{\partial\Omega} = p_0$ is a constant. Moreover, since temperatures drop from $\sim$ keV to $\sim$ eV in the edge, $p_0 \approx 0$ is a reasonably good approximation. This boundary in a vacuum-like region is well-motivated, but
only the gradient $\nabla p$ matters for force-balance. However, the
magnetic jump conditions in free boundary depend on 
$p_0$ through total-pressure balance.
Overall, this CBS is an advantageous choice because: 
\begin{itemize}[leftmargin=*]
    \item 
\textit{The closure becomes physically clean and robust.}
The problem is no longer choosing $\partial\Omega$ around a region where current and pressure can be changing rapidly. 
The pressure at the boundary, and the shape of the boundary itself need not be updated by a full transport solve when finding a self-consistent equilibrium (although presumably consistency must be enforced on the pressure profile in the volume).

Moreover, if one chooses $\partial \Omega$ outside the transition layer, where $J|_{\partial\Omega} = p_0 = 0$, then moving the computational boundary a bit farther outward, through a region that is already vacuum-like, leaves the form of the boundary conditions unchanged. This boundary may not be a flux surface, but it will be a $p_0 = 0$ isobar, so we address this important case later. The precise location of that boundary then becomes largely a gauge choice for the vacuum continuation (since between the first boundary and the second boundary there is only vacuum, which is uniquely determined in that toroidal shell), not a change in the interior plasma equilibrium.
\item \textit{The corresponding free-boundary problem becomes compatible with the fixed-boundary one.} The $J|_{\partial\Omega} = p_0 = 0$ assumptions potentially remove the additional conditions on free boundary problems since this is now a vacuum-continuation problem. If the tangential component of $B$ can be made continuous across the interface, there is no sheet current. Then the magnetic field is simply a continuous function across $\partial \Omega$. In other words, if a fixed-boundary solution is solved with this boundary condition, then the free-boundary problem supplies the vacuum continuation outside, but otherwise does not change the plasma solution. 
In particular, stellarator coil optimization is typically solved by reducing the $B\cdot n$ errors on a given computational boundary from a fixed-boundary equilibrium. With $J|_{\partial\Omega} = p_0 = 0$, minimizing $B\cdot n$ produces the true free boundary solution; otherwise, the coils only reproduce the fixed-boundary solution.
\item \textit{The closure is a natural choice for initializing time-dependent or finite-element MHD codes.} There are now time-dependent resistive MHD codes for stellarator geometries~\cite{nikulsin2022jorek3d,zhou2023m3dc1,patil2023updates}. Currently, they appear to be limited to ``fixed-boundary'' problems where they are initialized e.g. from a VMEC solution and only simulated within the pre-defined  boundary, ignoring the edge dynamics and transition layer. 

\end{itemize}



Clearly, this move to a surface with $J|_{\partial\Omega}=0$ and $p|_{\partial\Omega}=p_0$ is not a neutral rewrite of the original problem. Changing the boundary
changes the surface shape, enclosed volume and flux, and the exterior
vacuum field or coil system needed to support that new surface. In 3D it
may also change the admissible topological class if the edge
region contains islands or stochastic field lines. 

\section{The CBS and stage-two optimization}\label{vmec-and-desc}

\subsection{Resistive time-dependent and relaxation codes}\label{mrx}

The boundary condition $J \times n|_{\partial\Omega} = 0$ appears naturally in
resistive relaxation problems. In a resistive substep one solves an elliptic curl-curl problem for the magnetic update, and the weak form naturally produces a boundary term involving the tangential component of the current or, equivalently, the tangential electric field.

\paragraph{Boundary layer:}
For a simple resistive Ohm law $E=\eta J-u\times B$, a perfectly
conducting wall gives $n\times E|_{\partial\Omega}=0$; when applied to a static resistive substep, this reduces to the homogeneous natural condition $J \times n|_{\partial\Omega} = 0$ on the wall. A boundary layer of width $\sqrt{\eta \, dt}$ forms near the wall. We present a short illustration in Appendix~\ref{sec:boundary_layer}.
As $\eta \to 0$, this boundary layer collapses to a step function and to the best of our knowledge, most codes do not specify $J$ on the boundary any longer.

This is therefore a natural resistive wall closure for the time-dependent problem. 
If equilibrium force balance additionally holds on a flux-surface (perfectly conducting wall) or isobar boundary (a typical boundary condition for an edge layer) with nonzero tangential magnetic field, then the tangential part of $J\times B=\nabla p$ gives $J\cdot n|_{\partial\Omega}=0$; combined with $J\times n|_{\partial\Omega}=0$, this yields $J|_{\partial\Omega}=0$. Of course, force balance need not hold in time-dependent simulations.

The pressure is typically advanced using the second moment equation and generally taken with the natural choice of homogeneous Dirichlet boundary data. In that precise sense time-dependent resistive MHD codes are especially
well matched to the $J\times n|_{\partial\Omega} =0$, $p|_{\partial\Omega} = p_0$ boundary model.
\paragraph{Explicitly enforcing equilibrium at the boundary:}
Moreover, structure-preserving mixed finite-element codes~\cite{hu_stable_2017, gawlik_finite_2022, carlier_variational_2024, he_topology-preserving_2025}, including the recent 3D MHD non-nested equilibrium solver MRX~\cite{blickhan2025}, are naturally defined with $J \in H^{\mathrm{curl}}_0$. The final solution will satisfy force balance to high accuracy, so one may also impose the stronger condition $J|_{\partial\Omega} = 0$, but this will be a weakly enforced boundary condition (e.g. via a penalty method), reflecting the fact that this is a condition imposed by the equilibrium, not Maxwell's equations alone.
Note lastly that this outer boundary condition enters through the weak
formulation of the update equations themselves. That means that the non-nested equilibrium that magnetic relaxation or time-dependent codes find will \textit{fundamentally} depend on the choice of CBS. Indeed, for essentially any non-nested equilibrium code, the final equilibrium can depend on the update operator, the admissible variations, the invariants retained
by the algorithm, and the CBS used during
relaxation or continuation \cite{hirshman2011,peraza2017fb,peraza2017,blickhan2025,reiman1986,hudson2002,suzuki2006,schmitt2025}.

\subsection{Continuation of nested flux equilibria}\label{sec:continuation_nested_flux}
Given the differences in the CBS, there is an open question how to use time-dependent codes with resistive boundary conditions (or equilibrium relaxation codes like MRX) to evolve an initial nested-surface equilibrium solution. Nested flux surface codes generally do not assume profiles such that $J \times n|_{\partial\Omega} = 0$, and force balance may not be numerically well-satisfied so that $J\cdot n|_{\partial\Omega} = 0$ may not hold exactly on the boundary.
%
\paragraph{Fixed boundary traces:}
If the goal is to relax an existing fixed-boundary equilibrium to a nearby solution with islands and chaos, we propose that one imposes
\begin{align}
\label{eq:finite_traces}
    g_0 &:= (B_0\cdot n)|_{\partial\Omega} \in H^{-1/2}(\partial\Omega), \\ 
    \tau_0 &:= (J_0 \times n)|_{\partial\Omega} \in  H^{-1/2}(\dvg_{\partial \Omega}, \partial \Omega),
\end{align} 
at initialization, where 
\begin{multline}
    H^{-1/2}(\dvg_{\partial \Omega}, \partial \Omega) := \{ v \in H^{-1/2}(\partial\Omega; \mathbb{R}^3) : \\
    \dvg v \big |_{\partial \Omega} \in  H^{-1/2}(\partial\Omega; \mathbb{R}) \}.
\end{multline}

Here $B_0$ and $J_0$ are presumed time-independent and given from an initial condition, e.g. a fixed-boundary or free-boundary nested equilibrium solution. The new computational boundary can be specified interior or exterior to the previous boundary. If the new computational boundary is taken on the exterior, we can use virtual casing~\cite{hanson2015,toler2024direct} to evaluate the plasma contribution to $g_0$, and if current sources are specified, the vacuum contributions can be computed for $g_0$ by the Biot Savart law.

This boundary condition is kept constant in a numerical scheme by using structure-preserving finite-element methods for only the increment (which satisfies homogeneous boundary conditions), rather than on the total field. We have accounted for $g_0 \neq 0$, although this is zero for any existing equilibrium solvers since the boundary is assumed to be a flux surface. The fact that $B$ is divergence-free also imposes the compatibility condition
\begin{align}
\label{eq:consistency_condition}
\oint_{\partial\Omega} g_0 \, dS = 0.
\end{align}

\paragraph{Lifting:}
The full update becomes:
\begin{align}
\label{eq:incrementbcJ}
    B^{k+1} &= B^{k} + \delta B, \; J^{k+1} = J^{k} + \delta J,
    \intertext{in $\partial\Omega$ and }
    B^{k+1} \cdot n &= g_0, \; J^{k+1} \times n = \tau_0, \;
    \delta B \cdot n = \delta J \times n = 0, \notag
\end{align}
on $\partial\Omega$. We have that $\delta B \in \Hdivzero_0 \cap \Hcurl$ and $\delta J \in H^{\mathrm{curl}}_0(\Omega)$.
Thus the update rule is imposing on the total field the boundary tangential-current
trace inherited from the nested surface initial condition, while asking the \emph{update} $\delta J$ to satisfy the homogeneous wall condition. This is the usual \emph{lifting} approach of prescribing non-zero boundary conditions in finite element methods. 

In that sense the
CBS model is kept the same across the two codes: the initial condition on the boundary fixes the total trace, and the update rule is prevented from altering it at the boundary. Similarly one may hold the original boundary pressure value fixed by adding it to the pressure solve as a Dirichlet boundary condition, or in the case of MRX, the elliptic problem in Eq.~\eqref{eq:genericpelliptic}. This allows the further relaxation from fixed-boundary nested flux surface codes without immediately overwriting the boundary values inherited from the initial equilibrium. By contrast, enforcing $n\times J^{k+1}=0$ on the total field at every iteration would systematically pull the solution toward a different CBS model (and very possibly, a different 3D MHD equilibrium) by erasing the original CBS.

Moreover, in time-dependent codes, the subsequent dissipative evolution model has its
own wall conditions, source terms, and transport closure that must be
tuned if the imported equilibrium is to remain approximately stationary
\cite{zhou2021m3dc1,zhou2023m3dc1}. Part of the required tuning is in fact a \textit{consequence} of this mismatch between the CBS used in the initial condition and the CBS in the time-dependent code. 

\subsection{Consequences for stage-2 stellarator coil design}\label{consequences-for-stage-2-stellarator-coil-design}
The CBS choice has a serious consequence for the robustness of stage-two coil design for finite-$\beta$ stellarators. The reason is that if coils are optimized to \textit{perfectly} match a fixed-boundary surface, they will still not generally satisfy the full jump conditions required for a free-boundary solution. However, with the CBS $J|_{\partial\Omega}=p_0 = 0$, no change is required, since the fixed-boundary and free-boundary solutions will already be consistent; with this CBS, the only difference is that the boundary is free to move around in one of the formulations.

\paragraph{Mean-square normal-field error:}
The standard stage-2 stellarator coil objective penalizes the normal
field on a target plasma surface. In the two-step workflow, the
plasma-optimization stage first selects a fixed-boundary equilibrium,
after which the coil-design stage searches for external coils whose
vacuum field reproduces the required normal field on the target plasma
surface $\partial\Omega$ \cite{merkel1987,landreman2017,henneberg2021,ku2010,boozer2015}. Both winding surface methods~\cite{landreman2017,fu2025global,panici2025surface,fu2025flexible} and filamentary-coil methods~\cite{zhu2017new,landreman2021simsopt,hudson2018differentiating,kaptanoglu2025reactor,gil_augmented_2025,nie2025focus} target the
same geometric quantity,
\medmuskip=0mu
\begin{align}
\min_{\mathcal C}\; f_B(\mathcal C)
:= \frac{1}{2|\partial\Omega|}\oint_{\partial\Omega}
\left( (B_\text{coil}(\mathcal C) - B_{p}^\text{ext})\cdot n \right)^2\,dS,
\label{eq:quadraticflux}
\end{align}
\medmuskip=4mu
where $\mathcal C$ denotes the coil degrees of freedom. In finite-$\beta$ design, the required external normal field $B_{p}^\text{ext}$ is commonly obtained from virtual
casing, so that the coils reproduce the needed external support field
rather than zero normal field directly \cite{hanson2015,ku2010,conlin2024}. Practical
objectives then add regularization for engineering complexity, coil
length, curvature, separation, or current density, but the central
plasma-surface term remains the mean-square normal-field error
\cite{landreman2017,henneberg2021,jorge2024,gil_augmented_2025}.

\paragraph{Free-boundary objective:}
This objective is geometrically correct for reproducing a prescribed
magnetic surface, but it is weaker than reproducing the same
self-consistent free-boundary equilibrium. A natural stronger target is
therefore a \emph{free-boundary residual} on a candidate surface,
\medmuskip4mu
\begin{align}
\label{eq:freeboundaryobjective}
f_{\mathrm{FB}}(\mathcal C,\partial\Omega)
:= \oint_{\partial\Omega} \left( w_n |R_n|^2
+  w_t |R_t|^2
 + w_T |R_T|^2 \right) dS,
\end{align}
\medmuskip4mu
with $R_n$, $R_t$, and $R_T$ defined from the free-boundary jump and
traction conditions and design weights $w_{\{n,t,T\}}$. This notation $f_{\mathrm{FB}}(\mathcal C,\partial\Omega)$ implies that the full free-boundary problem is solved by letting the surface itself change. Generally the full free-boundary problem is fragile and complex to solve~\cite{constantin2022freeboundary,enciso2025jems,hudson2020specfb} although there is ongoing work to attempt to resolve this~\cite{conlin2024}; in practice, we can perform traditional stage-two optimization with fixed $\partial\Omega$, and instead ask that the coils are given enough degrees of freedom to match all of the conditions that are required for a true free boundary solution. 

\paragraph{Uniqueness:}
Moreover, with the vacuum-like $J|_{\partial\Omega}=p_0=0$ CBS, $R_t$ and $R_T$ reduce to the statement that the interior and exterior vacuum
continuations represent the same magnetic field on the interface. If the vacuum-field topology and global
flux or circulation data are fixed, uniqueness of the vacuum
continuation means that the only remaining design target is the
normal-field residual.
In other words, if the goal is the
usual one---reproduce the external support field of a fixed-boundary
equilibrium on a prescribed target surface---then the standard
normal-field objective \eqref{eq:quadraticflux} remains the right target.
If instead the goal is to create coils that will be self-consistent with the eventual
free-boundary calculation, then in some settings we should directly minimize the
corresponding free-boundary residuals.

\section{Specifying a general computational boundary}
\label{sec:general_boundary}


A traditional ideal fixed-boundary equilibrium solver prescribes a flux surface.
Moreover, profile data are then specified relative to that boundary. We have made the case that a good choice of the computational boundary uses $J|_{\partial\Omega}=p_0 = 0$. However, all CBSs, especially in 3D, suffer from a potential issue. What if there is no flux surface in the edge or vacuum region? What if the boundary lies directly on a rational surface, e.g. $\iota = 1$, where we know large magnetic islands will appear? 

\paragraph{Vacuum case:}
It is instructive to consider the vacuum case first. Suppose that some current sources (e.g. coils) are defined, and the Biot Savart law determines the magnetic field everywhere in space. In general, there will not be any simply-connected toroidal surface where $B\cdot n|_{\partial\Omega} = 0$. Instead, one can choose a bounded computational surface and evaluate the coil fields there, which prescribes some nontrivial boundary trace $g_0$ from Eq.~\eqref{eq:finite_traces}. This now gives a well-posed elliptic problem with inhomogeneous Neumann boundary condition in the interior $B = B_\mathfrak{H} + \nabla \phi$, where
\begin{align}
\label{eq:vacuum_neumann_inhomogeneous}
\nabla \times B_\mathfrak{H} = \nabla \cdot B_\mathfrak{H} &= 0, & \Delta \phi &= 0,        && \text{in }\Omega, \notag \\
                  B_\mathfrak{H} \cdot n      &= 0, & \nabla \phi \cdot n &= g_0, && \text{on }\partial\Omega,
\end{align}
where $B_\mathfrak{H}$ is the harmonic contribution fixed by the total toroidal flux $\Phi_0$, $g_0$ the boundary trace, and then $\phi\in \Hone(\Omega;\mathbb{R})$ will be uniquely determined by Eq.~\eqref{eq:vacuum_neumann_inhomogeneous} together with $\int_\Omega \phi \, dV = 0$.


\paragraph{Non-vacuum case:}
With finite pressure, the situation is more complex because we may want to specify some form of boundary conditions for $J$ and $p$. Since we have proposed the CBS using a vacuum-like region, this computational boundary would generically be a $p_0 = 0$ isobar but not a flux surface.
In fact, we will write down an algorithm for solving the equilibrium problem with a general computational boundary (not a flux surface or an isobar) in Appendix~\ref{sec:appendix_relaxation}. This  fixed-boundary finite-$\beta$ formulation with finite boundary traces is not standard.  Perhaps the only potential example is in HINT or HINT2, which sometimes works in real-space coordinates and uses a computational boundary with the normal component of the magnetic field fixed at the computational boundary~\cite{wiegmann2009hint2,suzuki2014hint2boundary}. Free-boundary SPEC calculations also introduce an auxiliary computational boundary outside the plasma target surface, and published examples note that this auxiliary boundary need not be a flux surface of the vacuum field~\cite{hudson2020specfb}. We emphasize that these previous examples use a free boundary.

We have already justified that a flux surface is also an isobar if force-balance holds, but the converse fails
in general. 
An isobar boundary can describe a broader class of computational domains than specifying a flux surface. It is naturally compatible with relaxed, resistive, or
increment-based formulations in which the boundary need not be a flux surface; Appendix~\ref{sec:appendix_relaxation} formulates the corresponding magnetic relaxation problem on any computational domain (not necessarily a flux surface or an isobar) with general finite normal traces in Eq.~\eqref{eq:finite_traces}.

\paragraph{Isobars through chaotic regions:}
We have also shown in Sec.~\ref{sec:force_balance_on_boundary} that away from the zero-sets of $g_0$, this requires $\partial_n p|_{\partial\Omega} = 0$ to satisfy force balance.
A constant-pressure, zero-pressure-gradient boundary is a good idealization for a surface chosen inside a pressure-flattened chaotic region; the field lines may densely fill a three-dimensional
volume or a substantial part of it~\cite{kanno2002,suzuki2020,enciso2025jems}. In such a flattened region, there is no magnetic flux surface but many closed surfaces bounding
$C^{1,1}$ toroidal solids can be treated as admissible isobaric computational boundaries. The choice of isobar is clearly not unique, and the solution will somewhat depend on this choice, as should be clear since each isobar typically describes a different interior volume. 



\section{Conclusion}\label{conclusion}
We have made the case that the computational boundary for 3D fixed-boundary MHD equilibrium calculations should, when the modeling goal permits it, be chosen in a vacuum-like part of the edge so that $p|_{\partial\Omega} = p_0$ and $J|_{\partial\Omega}= 0$; if the boundary is also intended to be flux surface, then $B\cdot n|_{\partial\Omega}= 0$ is imposed as usual. A weaker wall-style condition $J\times n|_{\partial\Omega}= 0$
implies $J|_{\partial\Omega} = 0$ only
when combined with the appropriate force-balance and
geometric hypotheses discussed above. Without this CBS, stage-two stellarator optimization should in principle directly target the
finite-$\beta$ \textit{free-boundary} conditions on a given candidate surface, not only the normal-field residual. With the vacuum-like closure and fixed global vacuum-field data, the matching conditions reduce to the vacuum continuation problem; in that special case the standard stage-two normal-field objective remains consistent with the free-boundary continuation.

Lastly, we have made the case that fixed-boundary calculations can, and in some cases \textit{should}, be formulated with a general computational boundary. The Appendix~\ref{sec:appendix_relaxation} provides a compatible magnetic-relaxation formulation for finding non-nested MHD equilibria from prescribed computational boundaries that are neither flux surfaces nor isobars. 

\section*{Acknowledgements}
This work was supported through grants from the Simons Foundation under award 560651.
We thank Wrick Sengupta, Elizabeth Paul, Sophia Henneberg, Matt Landreman, Michael Zarnstorff, Omar Maj and Florian Hindenlag for useful discussions. We acknowledge the use of generative AI tools for drafting and other edits.

\appendix
\section{Magnetic relaxation on any computational surface}
\label{sec:appendix_relaxation}
We have motivated the choice of a general computational boundary (e.g. an isobar instead of a flux surface) for performing MHD equilibrium calculations. To our knowledge, existing fixed-boundary 3D MHD equilibrium codes for magnetic confinement fusion design have not implemented this capability. Therefore, towards a numerical method that is compatible with this formulation, we now construct a fixed-domain magnetic relaxation problem in which the magnetic field and current may have nonzero boundary traces inherited from an initial condition. This is the general case where the boundary need not be an isobar or a flux surface.

This is a natural setting when one begins from a fixed-boundary
equilibrium computed on one surface and then wishes to continue or relax the solution
on a different prescribed surface without discarding the already-determined boundary
field. It is also the cleanest way to compare a traditional fixed-boundary equilibrium (which are always specified such that $g_0 = 0$ and often specified with $\tau_0 \neq 0$) to a relaxation scheme such as MRX: the inherited boundary traces are regarded as part of the background state, while the update is constrained to preserve the chosen CBS.

Now suppose we are given an initial
field $B_0 \in \Bspace$ with associated current $J_0\in \Hcurl(\Omega)$ and the boundary traces in Eq.~\eqref{eq:finite_traces}.
We now impose the update conditions from Equations~\eqref{eq:incrementbcJ}
for all iterates of the relaxation. 
%
This formulation is convenient for three reasons. First, it separates the inherited boundary information from the new relaxation.
The total field may already have nonzero normal flux or nonzero tangential current
trace on ${\partial\Omega}$, but the update need not alter them. Second, it puts the update in essentially the same function class as the usual
homogeneous-boundary fixed-domain problem. Third, it clarifies precisely what is meant when we adopt the CBS from an initial condition from a nested flux surface solution. That being said, the following derivations use a no-slip condition that require the increment $\delta B\cdot n|_{\partial\Omega} = 0$ but $\delta J\times n|_{\partial\Omega} = (\nabla \times \delta B)\times n|_{\partial\Omega}$ does not need to vanish, providing an outlet to relax to equilibria with different initial and final tangential currents. 
We now analyze how these inhomogeneous boundary conditions change the magnetic relaxation formulation of the MRX code, although this applies more broadly to relaxation methods.

\subsection{Monotonic decrease of the energy.}\label{sec:energy_decrease}
With these boundary conditions, we define the magnetic energy $\varepsilon_B: \Bspace \rightarrow \mathbb{R}$ of the field $B$ by
    \begin{align}
    \label{eq:energy}
        \varepsilon_B := \int_\Omega \frac{|B|^2}{2} \, d V = \frac{1}{2} \| B \|^2_{L^2(\Omega)}.
    \end{align} 
Since $B_0$ is fixed in time, we can take the variation and obtain,
\begin{equation}
\delta \varepsilon_B
=
\int_\Omega B\cdot \delta B\,dV =
\int_\Omega B\cdot \nabla\times (v\times B )\,dV,
\end{equation}
where $v$ denotes the ideal-relaxation velocity.
Using
\begin{align}
\int_\Omega B\cdot (\nabla\times E)\,dV
=
&\int_\Omega (\nabla\times B)\cdot E\,dV
\\ \notag &-
\oint_{\partial\Omega} (n\times B)\cdot E\,dS,
\end{align}
we obtain
\medmuskip2mu
\begin{align}
\delta \varepsilon_B
=
\int_\Omega J\cdot(v\times B)\,dV
-
\oint_{\partial\Omega} (n\times B)\cdot(v\times B)\,dS.
\end{align}
Since $J\cdot(v\times B)=-v\cdot(J\times B)$, this becomes
\begin{align}
\notag
\delta \varepsilon_B
=
-\int_\Omega v\cdot (J \times B)\,dV
 -
\oint_{\partial\Omega} (n\times B)\cdot(v\times B)\,dS.
\end{align}
The boundary term may be simplified by a vector identity,
\begin{align}
(n\times B)\cdot(v\times B)
&=
(n\cdot v)|B|^2-(B\cdot v)(B\cdot n).
\end{align}
Hence the energy variation $\delta \epsilon_B$ equals
\thickmuskip1mu
\medmuskip1mu
\begin{align}
\label{eq:appendix_excess_energy_balance_2}
-\int_\Omega v\cdot (J \times B)\,dV
-
\oint_{\partial\Omega} \Bigl[(n\cdot v) |B|^2 - g_0(v\cdot B)\Bigr]\,dS.
\end{align}
\thickmuskip4mu
\medmuskip4mu
We would like to choose the velocity field so that both boundary terms vanish. The traditional choice for magnetic relaxation is $v = J\times B - \nabla p$, with $p$ satisfying the elliptic solve in Eq.~\eqref{eq:genericpelliptic}.
%
The original formulation satisfies $v\cdot n|_{\partial\Omega} = 0$ pointwise and therefore eliminates the first boundary term. 
The second boundary term is generally indefinite.
We could project $v$ further by constraining to additionally eliminate the boundary term proportional to $g_0$. Unfortunately, this requires a projection operator with a nontrivial nullspace; minima of the energy therefore would no longer need to satisfy force balance in the volume.

We are left considering under what circumstances a stationary point of the energy will satisfy volumetric force balance. A clean way to avoid boundary terms and restore force balance is to choose the artificial relaxation velocity with a no-slip
boundary condition~\cite{brinkman1949,temam1977,girault1986,boffi2013,john_finite_2016}. Let $v\in\Noslipspace$, $\epsilon>0$, and define
\begin{align}
    a_\epsilon(v,w)
    &= \int_\Omega v\cdot w\,dV +\epsilon \int_\Omega \nabla v : \nabla w\,dV,
\label{eq:brinkman_bilinear}
\end{align}
\medmuskip=1mu
where $\int_\Omega \nabla v : \nabla w = \int_\Omega \big( (\nabla \times v) \cdot (\nabla \times w) + (\nabla \cdot v) (\nabla \cdot w) \big)$.
\medmuskip=4mu

Given $J \times B \in [H^{-1}(\Omega)]^3$, the no-slip Stokes--Brinkman mobility is the unique $v = \mathcal M_\epsilon (J \times B) \in \Noslipspace$ satisfying
\begin{align}
    a_\epsilon(v,w) = \int_\Omega (J \times B)\cdot w\,dV =: \mathcal{L}(w)
\label{eq:regularized_leray}
\end{align}
for all $w \in \Noslipspace$. Equivalently, there is a Lagrange multiplier $p\in L^2(\Omega)$, $(p, 1)_{L^2} = 0$ such that, in weak form,
\medmuskip2mu
\begin{align}
(v,w)_{L^2}+\epsilon(\nabla \times v,\nabla \times w)_{L^2}-(p,\nabla\cdot w)_{L^2}&=\mathcal{L}(w), \\
(q,\nabla\cdot v)_{L^2}&=0, \notag
\label{eq:stokes_brinkman_mixed}
\end{align}
\medmuskip4mu
for all $w\in (\Hcurl_0 \cap \Hdiv_0)(\Omega)$ and $q\in L^2(\Omega)$. In strong notation this is the projected equation
\begin{equation}
 v-\epsilon\Delta v+\nabla p=J \times B,\,\,\, \nabla\cdot v=0,\,\,\, v|_{\partial\Omega}=0,
\end{equation}
but no separate pointwise Neumann condition for $p$ should be imposed; as usual, the pressure is the Lagrange multiplier enforcing incompressibility. The no-slip condition eliminates all boundary terms in \eqref{eq:appendix_excess_energy_balance_2}, regardless of whether $g_0$ is zero or nonzero. Therefore
\begin{align}
\delta\varepsilon_B
&= -\int_\Omega v\cdot (J \times B)\,dV\notag\\
&= -a_\epsilon(v,v)\notag\\
&= -\left(\|v\|_{L^2(\Omega)}^2+\epsilon\|\nabla v\|_{L^2(\Omega)}^2\right) \notag\\
&= -\left(J \times B - \nabla p, v \right)_{L^2(\Omega)} \notag\\
&= -\left(J \times B - \nabla p, (1 - \varepsilon \Delta)^{-1} (J \times B - \nabla p) \right)_{L^2(\Omega)} \notag\\
&=: -\| J \times B - \nabla p \|^2_{H^{-1}_\varepsilon(\Omega)}
\leq0.
\label{eq:brinkman_energy_decay}
\end{align}
We used in the fourth equality the fact that $v$ is divergence-free and hence $L^2$-orthogonal to all gradient fields including $\nabla p$. As $(1 - \varepsilon \Delta)$ is a positive definite operator, $\| \dots \|^2_{H^{-1}_\varepsilon(\Omega)}$ is a norm. We conclude that zero-dissipation $\delta\varepsilon_B=0$ states of the no-slip mobility satisfy
\begin{equation}
J \times B = \nabla p,
\end{equation}
which is the desired volumetric force-balance condition. Notice that the pressure is a Lagrange multiplier that need not be the ``physical pressure'' until the end state, since then it must be the pressure that satisfies force balance. 

\subsection{Modified relative helicity balance.}

In non-contractible domains where the magnetic field has a non-zero normal boundary trace $g_0$, we introduce a static, curl-free reference field $B_{\text{ref}}$ such that $g_0 = B_{\text{ref}} \cdot n|_{\partial\Omega}$. The increment field is defined as $\tilde{B} = B - B_{\text{ref}}$, which satisfies $\tilde{B} \cdot n|_{\partial\Omega} = 0$. The fields are decomposed into their rotational and harmonic parts: $B = \nabla \times A + B_{\mathfrak{H}}$ and $B_{\text{ref}} = \nabla \times A_{\text{ref}} + B_{\text{ref}, \mathfrak{H}}$. Because $\tilde{B} \cdot n|_{\partial\Omega} = 0$, we can fix the gauge such that $\tilde{A} \times n|_{\partial\Omega} = (A - A_{\text{ref}}) \times n|_{\partial\Omega} = 0$. 

We define the modified relative helicity $\mathcal{H}$ to account for both the gauge-invariant relative helicity and the harmonic contributions:
\begin{align}
    \mathcal{H} := \int_\Omega (A + A_{\text{ref}}) \cdot \tilde{B} + \tilde A \cdot (B_{\mathfrak{H}} + B_{\text{ref}, \mathfrak{H}}) \, dV.
\end{align}
It holds that $\partial_t B = -\nabla \times E = \nabla \times (v \times B)$ and $\partial_t B_{\text{ref}} = 0 = \partial_t B_{\mathfrak{H}} = 0$. Consequently, $\partial_t \tilde{B} = -\nabla \times E$ and $\partial_t A = -E + \nabla \phi$ for some gauge function $\phi$. Evaluating the time derivative of the first term (the standard relative helicity $\mathcal{H}_{\text{rel}}$), we get:
\medmuskip0.9mu
\begin{align}
    \frac{d\mathcal{H}_{\text{rel}}}{dt} = \int_\Omega (-E + \nabla \phi) \cdot \tilde{B} - (A + A_{\text{ref}}) \cdot (\nabla \times E) \, dV.
\end{align}
\medmuskip4mu
Because $\tilde{B} \cdot n|_{\partial\Omega} = 0$, the scalar potential term $\int_\Omega \nabla \phi \cdot \tilde{B} \, dV$ vanishes via integration by parts. Integrating the $\nabla \times E$ term by parts yields:
\begin{multline}
    \frac{d\mathcal{H}_{\text{rel}}}{dt} = -\int_\Omega E \cdot \big( \tilde{B} + \nabla \times A + \nabla \times A_{\text{ref}} \big) \, dV \\
    + \oint_{\partial \Omega} ((A + A_{\text{ref}}) \times E) \cdot n \, dS.
\end{multline}
Substituting the curls gives $\tilde{B} + (B - B_{\mathfrak{H}}) + (B_{\text{ref}} - B_{\text{ref}, \mathfrak{H}}) = 2B - (B_{\mathfrak{H}} + B_{\text{ref}, \mathfrak{H}})$. The no-slip boundary condition, $v|_{\partial\Omega} = 0$ on $\partial \Omega$, enforces $E \times n|_{\partial\Omega} = (v \times B) \times n|_{\partial\Omega} = 0$ and nullifies the boundary integral. Since $E \cdot B = -(v \times B) \cdot B = 0$, the time evolution of the relative helicity becomes:
\begin{align}
    \frac{d\mathcal{H}_{\text{rel}}}{dt} =  \int_\Omega E \cdot (B_{\mathfrak{H}} + B_{\text{ref}, \mathfrak{H}}) \, dV.
\end{align}
Now we compute the time derivative of the harmonic correction term:
\begin{multline}
    \frac{d}{dt} \int_\Omega \tilde A \cdot (B_{\mathfrak{H}} + B_{\text{ref}, \mathfrak{H}}) \, dV \\
    = \int_\Omega (-E + \nabla \phi) \cdot (B_{\mathfrak{H}} + B_{\text{ref}, \mathfrak{H}}) \, dV.
\end{multline}
To be specific, we fix the gauge such that $\phi|_{\partial\Omega} = 0$, which eliminates the $\phi$ term appearing after an integration by parts. This is consistent with the earlier gauge choice since 
\begin{align}
    0= \partial_t (A- A_{\text{ref}}) \times n|_{\partial\Omega} = \underbrace{-E\times n |_{\partial\Omega}}_{= \, 0} + \nabla \phi \times n|_{\partial\Omega}.
\end{align}
Hence, $\nabla_{\partial\Omega}\phi = 0$, and $\phi |_{\partial\Omega} = c(t)$.
The scalar potential has residual additive gauge freedom. A gauge transformation $A\to A + \nabla \chi$  sends $\phi \to \phi + \partial_t\chi$. Taking $\partial_t\chi = -c(t)$ sets $\phi|_{\partial\Omega} = 0$. Therefore there is a gauge where are conditions on $A$ and $\phi$ are consistent.
Returning to the overall calculation, the remaining term cancels the leftover volume integral from $\frac{d}{dt}\mathcal{H}_{\text{rel}}$. Therefore,
\begin{align}
    \frac{d\mathcal{H}}{dt} = (1-1)\int_\Omega E \cdot (B_{\mathfrak{H}} +B_{\text{ref}, \mathfrak{H}}) \, dV = 0.
\end{align}

If, moreover, the electric field is given by the resistive MHD Ohm's law
$E = -v\times B + \eta J$,
then
\begin{multline}
\frac{d}{dt} \mathcal H
=
- 2 \eta \int_\Omega J\cdot B\,dV \\ 
+ \eta \oint_{\partial \Omega} (A+A_{\text{ref}}) \cdot (J \times n)\, dS.
\label{eq:appendix_relhelicity_balance_3}
\end{multline}
The boundary term can be shown to not be explicitly gauge-dependent since $\partial_t(B\cdot n)|_{\partial\Omega} = 0$.
Thus the relative helicity is exactly preserved under ideal increments, and its  dependence on resistivity is unchanged from the original setting.

In other words, \textit{magnetic relaxation a prescribed computational boundary with inhomogenous boundary traces can still be formulated so that the magnetic energy monotonically decreases, a critical point of the energy implies volumetric force balance, and the relative helicity is preserved.}

\section{Model closure with vanishing boundary current}
If the boundary is chosen in a vacuum or sufficiently far-edge region, it is natural to
impose the total trace $J|_{\partial\Omega} = 0$ (or, in a weak formulation, both $J\times n|_{\partial\Omega}= 0$ and $J\cdot n|_{\partial\Omega}= 0$)~\cite{xu2019vacuum,guazzotto2021part1,guazzotto2021part2,fitzpatrick2024inverse}.
In this case, even with nontrivial $g_0$, we can conclude $\nabla p|_{\partial\Omega} = 0$ from force balance. This is not directly enforced  by (the now homogeneous) Neumann problem in Eq.~\eqref{eq:genericpelliptic}.
Assuming that the iteration proceeds towards a state with volumetric force balance, the final boundary is an isobar, since the gradient must then vanish on $\partial\Omega$. However, during the iterations, the boundary need not be an isobar. If we use the traditional incompressible $v = J\times B - \nabla p$,  the second term in Eq.~\eqref{eq:appendix_excess_energy_balance_2} is not eliminated, since
\begin{align}
\int_{\partial\Omega} g_0(v\cdot B)\,dS
= -\int_{\partial\Omega}g_0 B\cdot\nabla p\,dS. 
\end{align}
only vanishes \textit{if} we reach force balance. This is no longer guaranteed since the boundary term is indefinite. Therefore, even in this limited condition where we know that the computational boundary will be an isobar at force balance and $g_0 \neq 0$, the no-slip mobility formulation is the safer formulation because it removes the
boundary terms and guarantees that force balance is reached.

\section{Boundary layers}\label{sec:boundary_layer}
We illustrate the effect of $\varepsilon$ in the operator $(1 - \varepsilon^2 \, \Delta)$ that appears in the artificial relaxation of $v$ as well as the resistive case for $B$ (where $\varepsilon^2 = \eta \, dt)$.

Consider the equation
\begin{align}
    v - \varepsilon^2 \, \Delta v = u
\end{align}
with homogeneous boundary conditions for simplicity and $\Delta v = \nabla (\nabla \cdot v) - \nabla \times (\nabla \times v)$.

This solution will have a boundary layer near $\partial \Omega$ where $v$ deviates from $u$ to hit zero on $\partial \Omega$, while the two are mostly unchanged in the interior. The characteristic length scale of this transition region is $\varepsilon$. This can be seen in the following one-dimensional toy problem: Consider $u(x) \equiv 1$ on $[0,1]$. The equation
\begin{align}
    (1 - \varepsilon^2 \partial_x^2) v(x) = u(x) = 1 \quad \forall x \in (0,1)
\end{align}
with $v(0) = v(1) = 0$ has the solution
\begin{align}
    v(x) = 1 - \cosh((x-1/2)/\varepsilon) / \cosh(1/2\varepsilon)
\end{align}
or, to very good approximation as $\varepsilon \ll 1$,
\begin{align}
    u(x) - v(x) = e^{-x/\varepsilon} + e^{(x-1)/\varepsilon} + \mathcal O(e^{-1/\varepsilon}).
\end{align}

\bibliographystyle{apsrev4-2}
\bibliography{edge_closures_combined}

\end{document}